\documentclass[aps,prc,preprint,tightenlines,showpacs,preprintnumbers,amsmath,amssymb,superscriptaddress]{revtex4}

\usepackage{psfig}
\usepackage{graphicx}
\usepackage{dcolumn}
\usepackage{bm}

\begin{document}
%
\title{Pseudovector vs. pseudoscalar coupling in one-boson
exchange NN potentials}

\author{G. Caia}
\affiliation{Department of Physics and Astronomy, Ohio University,
Athens, OH 54701, U.S.A.}
\author{J.W. Durso}
\affiliation{Institut f\"ur Kernphysik (Theorie), Forschungszentrum J\"ulich,
D-52425 J\"ulich, Germany}
\author{Ch. Elster}
\affiliation{Department of Physics and Astronomy, Ohio University,
Athens, OH 54701, U.S.A.}
\affiliation{Institut f\"ur Kernphysik (Theorie), Forschungszentrum J\"ulich,
D-52425 J\"ulich, Germany}
\author{J. Haidenbauer}
\author{A. Sibirtsev}
\author{J. Speth}
\affiliation{Institut f\"ur Kernphysik (Theorie), Forschungszentrum J\"ulich,
D-52425 J\"ulich, Germany}

\begin{abstract}

We examine the effects of pseudoscalar and pseudovector coupling
of the $\pi$ and $\eta$ mesons in one-boson exchange models of the
$NN$ interaction using two approaches: time-ordered perturbation
theory unitarized with the relativistic Lippmann-Schwinger equation, 
and a reduced Bethe-Salpeter equation approach using the Thompson
equation.  Contact terms in the one-boson exchange amplitudes in
time-ordered perturbation theory lead naturally to the
introduction of $s$-channel nucleonic cutoffs for the interaction,
which strongly suppresses the far off-shell behavior of the
amplitudes in both approaches.  Differences between the resulting
$NN$ predictions of the various models are found to be small, and
particularly so when coupling constants of the other mesons are
readjusted within reasonable limits.
\end{abstract}


\maketitle

\section{Introduction}

Since the discovery and identification of the
pion~\cite{lop47,marbe47} as the strongly interacting meson
anticipated by Yukawa~\cite{yuk35} in 1935, most theoretical
efforts to construct a quantitatively accurate model of the
nucleon-nucleon ($NN$) interaction have used the pion-nucleon interaction
as the first building block.  Indeed, phase shift analyses of
$NN$ scattering data since 1959~\cite{cmms59} use the
one-pion exchange amplitude to fix the phase shifts of the high
orbital angular momentum partial waves, which are not individually
adjusted to fit the data.

Almost at the beginning of these efforts the question arose of
whether the fundamental coupling of the pion to the nucleon is of
the pseudovector or pseudoscalar type.  The question arises, of
course, from the fact that the fully on-shell one-pion exchange
amplitude derived from a $\pi NN$ interaction Lagrangian with
pseudovector coupling is identical in form to that from one with
pseudoscalar coupling.  In early attempts to go beyond the
one-pion exchange to the two-pion exchange~\cite{bw53,tmo},
pseudoscalar coupling appeared to demand suppression of ``pair
terms''---terms describing the contribution from intermediate
states with one or more antinucleons.  For the exchange of pions
with low momenta, this effectively reduced to pseudovector
coupling, for which the renormalizability of the theory is
doubtful, at best~\cite{serot82,bentz97}. Pion-nucleon scattering,
through the smallness of the scattering lengths, also strongly
suggested pseudovector coupling. Dispersion-theoretic results for
the two-pion exchange contribution~\cite{paris75,jrv75} based on
unitarity and analytic continuation of $\pi N$ amplitudes to the
$\pi\pi \rightleftharpoons N\overline N$ channel also implicitly
favored pseudovector coupling.

We realize now that the meson theory of nuclear forces is not a
fundamental theory but, at best, an effective theory.  Thus lack
of renormalizability in the usual sense is not a relevant
criterion for the rejection of one form of coupling or another.
Furthermore, the approximate chiral symmetry exhibited by QCD and
implemented with considerable success in chiral perturbation
theory ($\chi$PT)~\cite{Bernard97}, makes it clear that the
effective coupling of pions to nucleons, at least at low energies,
is pseudovector in character.  In other phenomena the picture is
more complex.  In pion electro- and photoproduction analyses at
low energies pseudovector coupling is preferred, whereas at higher
energies pseudoscalar coupling provides a more economical
description~\cite{Drechsel92,Drechsel99}.   In the light of this
evidence it is clear that any of the $NN$ interaction
models claiming to be realistic should, to some degree, include pion
exchange with pseudovector coupling.

With the exception of some recent models based on baryon
$\chi$PT~\cite{ordon94,ordonez,epelbaum98,epelbaum,entem}, $NN$
models of the past three decades include, in addition to the
one-pion exchange, contributions due to exchange of heavier
mesons, whether explicitly, as in the various one-boson exchange
(OBE)
models~\cite{rmpobe,ptpobe,signell,erkelenz,kotthoff,hol75a,hol75b,mach89,nijm94,mach01},
or implicitly, as resonant $t$-channel exchange of two pions, as
in the dispersion-theoretic approaches~\cite{paris75,jrv75}.  Even
within the OBE models there is no single preferred approach. There
are those models that are based on a Bethe-Salpeter~\cite{bs51}
approach and in which the unitarizing equation is some
three-dimensional reduction of the Bethe-Salpeter
equation~\cite{bbs66,thom,erkelenz,hol75a,hol75b,mach89,bm90}, and
there are those, including some of the various Bonn
potentials~\cite{kotthoff,ehm87,mach89}, that are based on
time-ordered perturbation theory (TOPT). Our main interest in this
work is in TOPT, but we shall also examine differences between
pseudoscalar and pseudovector coupling in both of these
approaches. For that purpose we will utilize a specific
three-dimensional reduction of the Bethe-Salpeter equation known
as the Thompson equation \cite{thom}.

In covariant perturbation theory one starts with a Lorentz-invariant Lagrangian
density, from which one derives the Hamiltonian density and, from that, the
Hamiltonian.  For Lagrangians with scalar or pseudoscalar mesons without
derivative coupling, the interaction part of the Hamiltonian density is just
the negative of the interaction part of the Lagrangian density.  For
Lagrangians with derivative coupling or with vector mesons, however,
non-covariant ``contact'' terms arise in the Hamiltonian density.  These terms
are necessary to cancel the non-covariant terms in the meson propagators so
that, in any order of perturbation theory, the resulting amplitude is
covariant~\cite{weinberg}.  From a procedural point of view, this means that in the
Feynman rules one simply drops the contact terms and the non-covariant parts
of the propagators.

In TOPT, however, one does not use particle propagators.  They are
effectively supplied by the vertex functions and energy
denominators in the time-ordered diagrams.  In order to obtain
covariant results in TOPT starting from a Lagrangian density with
derivative coupling, or with vector mesons, one must include the
contributions  of the contact interactions in the Hamiltonian in
the appropriate order of the perturbation expansion.  Therefore,
for single pion exchange with pseudovector coupling in $NN$
scattering, i.e., in second order in the coupling constant, one
must include not just the meson exchange diagrams, but also the
four-point $NN$ contact interaction, as shown in
Fig.~1, and similarly for vector meson exchange.  Only then will
the result agree with covariant perturbation theory when all
external particles are on their mass shells.

The main focus of the work that follows is to compare the results of
inclusion of the full pseudovector-coupled pion exchange with
those of pseudoscalar coupling in one-boson exchange models based
on time-ordered perturbation theory.  This is not simply a moot
point: the pion-nucleon interaction used in
Ref.~\onlinecite{ehm87}, although nominally of pseudovector type,
is in fact pseudoscalar, and thus has a very different off-energy
shell behavior from pseudovector coupling. Furthermore, contact
terms in the vector meson exchange contributions, as well as gauge
terms, which have heretofore been ignored, will be retained.

We shall, in this work, restrict ourselves to examining one-boson
exchange models, since the contact terms in higher orders of TOPT
present additional difficulties that are not easily resolved.  We
will, as in all models of this kind, need to introduce cutoffs in
order to ensure convergence of the integral equations used to
unitarize the scattering amplitude. We will also compare the
results to those obtained from a reduced Bethe-Salpeter approach,
in which the problem of contact terms doesn't arise.

In Sect.~II we describe the models that we employ in the present
study. In particular, we give the equations that are used for
unitarizing the scattering amplitude and we discuss some specific
aspects of the vertex form factors. Finally, we outline the
strategy that was followed for fixing the free parameters of the
models. Results for phase shifts as well as for the deuteron
properties are presented in Sect.~III. We analyze the consequences
of pseudovector versus pseudoscalar coupling in TOPT as well as in
a model based on the Thompson equation and we also compare the
different approaches. In Sect. IV we summarize our results and
draw some conclusions. Technical details such as the underlying
Lagrangians, and the potential matrix elements in TOPT and for the
Thompson equation are summarized in the Appendices A, B, and C.

\section{The models}

In order to make the comparison of different treatments of
pseudoscalar meson exchange explicit we will focus on four
one-boson exchange models of the $NN$ interaction.  The first
model that we will consider is that of pseudovector coupling of
the pseudoscalar mesons, $\pi$ and $\eta$, in TOPT. The second
will be the same with the exception that pseudoscalar coupling of
the pseudoscalar mesons will be used. For the third and fourth
models we make the same comparison of pseudovector and
pseudoscalar coupling for the pseudoscalar mesons, but within a
Bethe-Salpeter approach to the problem using the Thompson
equation~\cite{thom} to unitarize the scattering amplitude.

The roster of exchanged mesons in each of the four models is
identical: pseudoscalar mesons $\pi$ and $\eta$, vector mesons
$\rho$ and $\omega$, and scalar mesons $\sigma$ and $a_0$.
Contact terms as well as gauge terms arising in the polarization sums
in vector meson exchange in TOPT will be retained.  In the reduced
Bethe-Salpeter approach, gauge terms in the vector meson
propagators will also be retained. The masses and quantum
numbers of the mesons are given in Table~\ref{props}.

The interaction Lagrangian densities for all of the meson-baryon
interactions in our model calculations are given in Appendix A,
along with the corresponding Hamiltonian densities. The matrix
elements of the corresponding second order potentials for the
TOPT-based models are presented in Appendix B. Shown schematically
in Fig. 2 are the interaction and the kinematics for the
potential, $V$, in the $NN$ center-of-mass frame. For this we
unitarize the scattering amplitude through use of the
Lippmann-Schwinger equation,
\begin{equation}
{\bf T} \ = \  {\bf V}\ +\ {\bf V}\frac{1}{W\ - \ {\bf H}_0\ +i\epsilon}{\bf T}
\end{equation}
or, more precisely,
\begin{widetext}
\begin{eqnarray}
<\vec{p}\ '\lambda_1'\lambda_2'\mid T(W)  \mid\vec{p}\lambda_1\lambda_2>&=&
<\vec{p}\ '\lambda_1'\lambda_2'\mid V(W) \mid\vec{p},\lambda_1\lambda_2>\ \nonumber\\
& + & \sum_{\mu_1 \mu_2}\int \frac{d\,^3q}{W-2E_q+i\epsilon}<\vec{p}\ '\lambda_1'
\lambda_2'\mid V(W) \mid\vec{q}\mu_1 \mu_2> \nonumber\\
& \cdot & <\vec{q} \mu_1 \mu_2\mid T(W) \mid\vec{p} \lambda_1 \lambda_2>.
\label{eq:2}
\end{eqnarray}
Here $W$ represents the total energy of the $NN$ system.

The Thompson equation, which we use to unitarize the scattering amplitude in the
other two models, is given by
\begin{eqnarray}
<\vec{p}\ '\lambda_1'\lambda_2'\mid {\cal T}(W)  \mid\vec{p}\lambda_1\lambda_2>&=&
<\vec{p}\ '\lambda_1'\lambda_2'\mid {\cal V}(W) \mid\vec{p},\lambda_1\lambda_2>\ \nonumber\\
& + & \sum_{\mu_1 \mu_2}\int \frac{d\,^3q}{W-2E_q+i\epsilon}<\vec{p}\ '\lambda_1'
\lambda_2'\mid {\cal V}(W) \mid\vec{q}\mu_1 \mu_2> \nonumber\\
& \cdot & \frac{m_n^2}{E_q^2}<\vec{q} \mu_1 \mu_2\mid {\cal T}(W)
\mid\vec{p} \lambda_1 \lambda_2>.
\end{eqnarray}
\end{widetext}
The simple prescription for obtaining the matrix elements ${\cal V}$
from the matrix elements $V(W)$ of Eq.~(\ref{eq:2})
is given in Appendix C.

Except for the cutoff functions $ F_j(W,\vec{p}\,',\vec{p}\,)$
that multiply the field-theoretic meson exchange amplitudes in the
potential matrix elements $V$ and $\cal V$, the input to our model
calculations is completely specified.  The cutoff functions, as
already mentioned, are needed for the convergence of the
scattering equation, but the form that one chooses is largely
arbitrary. One commonly used, especially in one-boson exchange
models, is the so-called ``multipole'' form,
\begin{equation}
\label{multipole}
 F_j(W,\vec{p}\,',\vec{p}\,) = \left[ \frac{\Lambda_j^2-m_j^2}{\Lambda_j^2
+(\vec{p}\,'-\vec{p}\,)^2} \right]^{n_j},
\end{equation}
where $\Lambda_j$ and $n_j$ are the free parameters and $m_j$
stands for the meson mass. Part of the appeal of this particular
cutoff is that the connection between the range at which the
cutoff becomes effective for that meson potential is simply
related to the cutoff parameter: $R_{j,{\rm cutoff}} \approx
\sqrt{n_j}/\Lambda_j$. Considered as a product of two mesonic form
factors, this is then interpreted as, or assumed to be, a
reflection of the effective size of the meson cloud in that part
of the $NN$ interaction.

In treating contact terms in the interaction, however, this form
immediately raises some difficulties.  The first is that, through
the dependence on the three-momentum transfer in the denominator,
it introduces effects of the contact terms into {\em all} angular
momentum partial waves, whereas the contact terms alone, which are
polynomials of low degree in the sine or cosine of the c.m.
scattering angle, contribute only to a few of the lowest partial
wave amplitudes. The contact term arising from pseudovector pion
coupling, for example, contributes only to $s$- and $p$-waves. In
adopting the multipole form, one singles out the three-momentum
transfer as the variable in which to cut off the potential, but
there is no compelling reason to do so, and good reason not to.
Indeed, in the effective field theoretic approach of Epelbaum et
al.~\cite{epelbaum}, a purely $s$-channel cutoff is used.

On the other hand, we wish to have the comparison presented in
this work make some contact with models in the
literature~\cite{ehm87,mach89,nijm94} that employ $t$-channel
cutoffs. Furthermore, we wish to apply the same cutoff to both the
meson exchange terms {\em and} the contact terms in order that one
term not receive excessive weight from one kinematic region in the
range of the loop integral as compared with the other. This is a
problem of long standing and we will not attempt to address it
here. Rather, we will simply adopt {\em ad hoc} the form
\begin{widetext}
\begin{equation}
\label{formfac}
 F_j(W,\vec{p}\,',\vec{p}\,) = \left[ \frac{\Lambda_j^2-m_j^2}{\Lambda_j^2
+(\vec{p}\,'-\vec{p}\,)^2} \right]^{n_j}
\left(\frac{\Lambda_N^4}{\Lambda_N^4+(W^2/4-E_p^2)^2}\right)^2
\left(\frac{\Lambda_N^4}{\Lambda_N^4+(W^2/4-E_{p'}^2)^2}\right)^2;
\end{equation}
\end{widetext}
that is, we take the form of Eq.~(\ref{multipole}) and multiply it
by a factor $\Lambda_N^4/(\Lambda_N^4+(W^2/4-E_{q}^2)^2)$ for each
nucleon line with momentum $\vec{q}\,$ entering or leaving the
interaction.  In all cases except as noted in Table~\ref{params}
below we take $n_j=2$.

This form provides the potential with both $t$- and $s$-channel
cutoffs.  Although it does not remove the objection, raised above,
of mixing contact terms into higher partial waves, we shall
mitigate that effect via the proviso that the parameter
$\Lambda_j$ for meson terms whose interactions include contact
terms be chosen large enough so as to have negligible effects in
partial waves with orbital angular momentum $l\ \ge \ 2$\,, except
for $^3D_1$ because of its coupling to $^3S_1$.

As there are free and nearly-free parameters in the models, a
simple comparison of the $NN$ phase shifts produced by the four
models for a given set of parameters would not, in our opinion, be
a useful way to present the results of our investigation.  Such a
comparison might be seen to favor one model over another, which is
not our intention.

Instead, for each of the four models, we perform a constrained
least-squares fit of the adjustable parameters to the $NN$ phase
shifts in the range of laboratory kinetic energies from 20 to
300~MeV taken from  the energy independent $NN$ phase shift
analysis SP40 of Ref.~\onlinecite{said}. By ``constrained'' we
mean that the parameters that are allowed to vary, such as
coupling constants and cutoff masses, are restricted to a range
that is consistent with values used in the various meson-exchange
models of the $NN$ interaction in the literature. The results of
these best fits will then be compared in detail. We will also, for
the sake of completeness,  make a comparison of pseudovector and
pseudoscalar coupling in the TOPT models with identical parameter
sets.

Admittedly, this is not a rigorous procedure, but it reflects
better the intention of our work.  We wish to examine whether the
terms that, in principle, should be included in one-boson exchange
models based on TOPT but have heretofore been omitted require a
major reworking of previous models, or if their effects can be
compensated for by relatively small adjustments in the parameters
of the other models.  Therefore we are not concerned with ``high
precision'' fits of the models to the phase shifts, but with
qualitatively acceptable fits of the magnitudes and energy
dependencies of the model results to the data, since further
refinements of the models, such as the inclusion of two-pion
exchange or effects of baryon resonances~\cite{ehm87}, would
necessitate refitting of the parameters and, presumably, result in
quantitatively better fits.

At this point we must inject a word of caution for the reader.  As
our models have both $s$- and $t$-channel cutoffs to regulate the
integral equations, one should not expect the cutoff masses that
we use, especially those for the mesonic form factors, to agree
well with those in the literature for models using different
cutoff schemes. The way that cutoffs are implemented in any model
is part of the model and thus have a large influence on the values
of the model's parameters.

Our procedure is first to select the parameters in the models that
we wish to vary and then set the limits of variation of each of
these parameter to values that we consider reasonable. For example
the $\pi N$ coupling constant, $g_{\pi}^2/4\pi$ is fixed at 13.8,
whereas the $\eta N$ coupling constant, $g_{\eta}^2/4\pi$, since
it is less well-known, is allowed to vary from  0 to 6.  Cutoff
masses, $\Lambda_j$, are allowed a fairly large range, but are
required to be greater than 1~GeV.  The meson masses are fixed at
the values given in Table~\ref{props}. For the nucleonic cutoff
mass, $\Lambda_N$\,, we explored a range of values and found that
good fits with all the models could be obtained for $\Lambda_N$\
between 600 and 900~MeV. For values below 600~MeV the potential
was too strongly suppressed and for values above 1~GeV the cutoff
had almost no effect.  We therefore fixed the value of
$\Lambda_N$\ at 700~MeV in all the models.  With this choice the
contributions to the scattering amplitude at low c.m. energies
from intermediate states with energies above the pion production
threshold are strongly suppressed. (We should remark here that
this value of $\Lambda_N$\ is of the same order of magnitude as
the $s$-channel cutoff employed by Epelbaum et al.~\cite{epelbaum}
in their effective field-theoretic approach.) We then perform a
least squares search on the variable parameters.

\section{Results}

The parameters for our four models are shown in
Table~\ref{params}, where we give the complete set of coupling
constants and cutoff masses for the TOPT models, and in
Table~\ref{tparams}, where we show them for the models based on
the Thompson equation~\cite{thom}.  The phase shifts predicted by
the various models we have considered are shown in Fig.~3, where
we present the results for the TOPT models, and in Fig.~4, where
we present them for the Thompson equation models and, for purposes
of comparison, also for the TOPT model with pseudovector coupling.
Phase shifts are shown for partial waves only up to $J=3$,
omitting $\epsilon_3$, since the differences between the phase
shifts calculated with the various models are almost imperceptible
in the omitted phase shifts. Since the pion coupling is the same in
all models, it is clear that the higher partial waves in the
energy regime considered are practically identical. The deuteron
properties calculated from the models are given in Table IV.

As our primary interest is in the differences due to alternative
couplings of the pseudoscalar mesons, we shall first examine the
results for the TOPT models.  We shall then turn our attention to
the results of Thompson equation models.  Afterwards, we shall
briefly compare the results of the models in the two approaches.

\subsection{TOPT models}

The two ``best fit'' TOPT models---curves PV and PS in
Fig.~3---both give reasonably good descriptions of the phase shift
data. Where there are discernible differences, it is difficult to
form a consistent picture of the effects of the two alternative
couplings that cannot be compensated by relatively small
adjustments in the cutoffs or in the coupling constants of the
other mesons.  As mentioned before, the pion coupling constant is
the same in all models. In order to demonstrate the difference of
the two coupling schemes as it arises when there is no
readjustment of the other meson parameters, Fig.~3 includes a
calculation, labelled PS0, in which the only difference from model
PV is the change of the coupling of the $\pi$ and $\eta$ mesons
from pseudovector to pseudoscalar. A comparison of the results of
PV with PS0 shows that, even with no readjustments of the cutoffs
or coupling constants of the other mesons, the differences between
pseudovector and pseudoscalar coupling for $\pi$ and $\eta$ is
quite small, especially at low energy.  There is, however, one
glaring exception: the predictions for the mixing parameter
$\epsilon_1$. This parameter as calculated from PS0 is almost
double the value from PV---and the data--- throughout the energy
range shown, thus indicating that the tensor force resulting from
the two coupling schemes is quite different.

Indeed, it is the need to describe $\epsilon_1$ more accurately
that largely drives the changes in the coupling constants and
cutoffs in model PS from model PV. With very few exceptions, the
results of PS0 are closer than those of PS to the results of PV.
The exceptions, as one might expect from their connection with
$\epsilon_1$, are the phase shifts $^3{\rm S}_1$ and $^3{\rm
D}_1$, to which $\epsilon_1$ is coupled.

Comparing the coupling constants and cutoffs of PV and PS, one
sees that the $\eta$ is essentially unchanged.  The $\rho$
coupling in PS is slightly smaller, but that is partially offset
by a cutoff mass that is slightly larger.  The greatest
differences between the parameters in the two models are in the
$\sigma$ and $\omega$.  The $\omega$ coupling in PS is about 25
percent larger than in PV, although its cutoff mass is smaller,
which tends to compensate for the increased repulsive strength at
short distance.  The increase in the $\sigma$ coupling is
necessary, apparently, to provide attraction at intermediate range
to counter the greater repulsion due to the $\omega$.

While the picture isn't entirely clear, the competition between
increased attraction due to the larger $\sigma$ contribution at
intermediate range and increased repulsion at short range due to
the larger $\omega$ contribution appears evident.  In the
mid-peripheral uncoupled phases where differences can be observed,
i.e. $^1{\rm D}_2$ and $^3{\rm D}_2$, PS is slightly more
repulsive than PV, while in the more peripheral phase shift
$^1{\rm F}_3$, the reverse is true.  For higher partial waves,
which are not shown in the figures, the phases are almost entirely
given by the on-shell pion exchange term, but the effect is, as
one would expect, the same as in $^1{\rm F}_3$, although the
differences are so small as to be invisible on graphs.

\subsection{Thompson equation models} 
The phase shift fits for the two models that employ the Thompson
equation are shown in Fig.~4, along with the results for the
TOPT-based model PV discussed in the previous section.  The curves
labelled TPV show the results for the model with pseudovector
coupling of the $\pi$ and $\eta$ mesons, while those labelled TPS
show the results for the model with pseudoscalar coupling of the
$\pi$ and $\eta$ mesons.  Both TPV and TPS are best-fit results in
the restricted sense discussed in Section~II, in which bounds are
placed on the range of the adjustable parameters.

As in the case of the TOPT models, the phase parameter that shows
the most striking difference between pseudovector and pseudoscalar
coupling is the mixing parameter $\epsilon_1$.  Using the models
with the Thompson equation with the cutoffs implemented as we have
described, it is impossible to achieve a satisfactory description
of $\epsilon_1$ while simultaneously keeping the coupling
constants within reasonable bounds when pseudoscalar coupling is
used for the $\pi$ and $\eta$ mesons.

Apart from $\epsilon_1$, the only other phase shifts that reveal
any noticeable difference between pseudovector and pseudoscalar
coupling in the Thompson equations models are $^1{\rm P}_1$,
$\epsilon_2$, and $^3{\rm D}_3$, and even there the differences
are rather small. For the most part, the two Thompson equation
results are closer to each other than either is to PV or to PS.
This result isn't surprising, since most of the parameters in TPV
and TPS are the same, which reflects the fact that they are at the
limits of their permitted ranges. An interesting result is that
the $t$-channel cutoff of the $\sigma$ meson, $\Lambda_\sigma$, is
extremely large---10~GeV, which is effectively infinite.  This
suggests that the $s$-channel cutoff has a very powerful effect in
the Thompson equation models, and the large value of
$\Lambda_\sigma$ reflects an effort of the fitting program to
increase the attractive effect of the $\sigma$ exchange
contribution to counter the increased strength of the $\omega$
contribution. Indeed, with very few exceptions, both of the
Thompson models are more repulsive than PV, as one might expect
from the large value of $g_\omega^2$.

A slightly different view of the models that we have considered is
provided by the deuteron parameters compiled in
Table~\ref{deuteron}, which shows some small---but consistent---
differences between them. All four models are adjusted to fit the
deuteron binding energy very accurately.  Both TOPT models give a
very good value of the quadrupole moment with a relatively low
$d$-state probability, which is characteristic of TOPT models,
although the tendency of the Thompson models to have relatively
large $d$-state probability is mitigated here, presumably because
of the strong $s$-channel cutoff that was not present in earlier
work~\cite{mach89}. In both the TOPT and Thompson equation models,
pseudoscalar coupling results in a lower $d$-state probability
than pseudovector coupling.  The asymptotic $s$-wave, $A_S$, is
somewhat high for the TOPT models, although the asymptotic
$d$-to-$s$ ratio, $A_D/A_S$, is acceptable.  As expected, the
$d$-state probability of the Thompson models is larger than in the
corresponding TOPT models, resulting in a lower value of $A_S$.
The value of $A_D/A_S$, however, is not very different among the
four models, which probably explains the smaller quadrupole
moments for the Thompson equation models: the slightly greater
$d$-state probability in the Thompson equation deuteron wave
functions is not sufficient to counter the  slightly more compact
structure of the deuteron that the Thompson equation produces.

\section{Summary and conclusions}

It is clearly impossible to draw any general conclusions
concerning the effects of pseudoscalar as opposed to pseudovector
coupling of pseudoscalar mesons in the $NN$ system. Any
statements that we make are necessarily qualified by their model
dependence.  It is nevertheless useful to summarize our approach
and findings and to note, if possible, any tendencies within the
limited context of the models that we have investigated.

In the first place, we wished to treat pseudovector coupling
properly in the TOPT approach, using the Lippmann-Schwinger
equation to generate unitary scattering amplitudes from the
one-meson exchange amplitudes, and to compare it with pseudoscalar
coupling. The presence of contact terms in the pseudoscalar meson
exchange terms, as well as in the vector meson exchange terms with
tensor coupling, led us to introduce nucleonic---i.e.
$s$-channel---form factors in addition to the $t$-channel form
factors that are usually employed in meson exchange modes of the
$NN$ interaction.

For the purpose of comparison with a different approach, we
examined the difference between the two coupling schemes in the
context of a model based on a particular version of
three-dimensional reduction of the Bethe-Salpeter equation, the
Thompson equation.  In this approach the energy denominators and
the off-shell continuations of the meson exchange amplitudes
differ from those of TOPT.  In particular, no contact terms
appear.  In order to keep the comparison as close as possible, we
chose similar form factors to those in the TOPT-based models that
we studied.

Within each model we made a restricted best fit to the $NN$ data,
allowing meson coupling constants and form factors to vary within
broad limits chosen with regard to values of these parameters
found in earlier works.  We also examined the effect of simply
changing the pseudovector coupling of the $\pi$ and $\eta$ to
pseudoscalar, within the TOPT approach.

Not surprisingly, perhaps, most of the differences in the
scattering and bound state properties calculated with the various
models were quite small, which probably reflects the fact that the
cutoffs were strong enough to strongly suppress differences in the
off-shell behavior of the two coupling schemes.  The most dramatic
difference in the TOPT models appeared in the mixing parameter
$\epsilon_1$ when the simple change of pseudovector to
pseudoscalar coupling was made. This should not be surprising,
since the effect of the contact terms is limited to $s$- and
$p$-waves and states coupled to them. The diagonal pion exchange
amplitude is rather weak so that it is in the relatively small
mixing parameter $\epsilon_1$, which is dominated by $\pi$ and
$\rho$ exchange, that changes in the short-range behavior of the
tensor force is most strongly felt.  Indeed, most of the
differences between the ``best fit'' pseudovector and pseudoscalar
models was due to the readjustment of parameters in the latter
needed to produce a better fit to $\epsilon_1$.

The same effect was observed in the Thompson equation models, with
the largest relative difference between the pseudovector and
pseudoscalar coupling versions remaining after refitting coupling
constants appearing in $\epsilon_1$.  Other differences between
the two Thompson equation models were very small, owing in large
part to the fact that several of the adjustable parameters were at
the limits of their permitted ranges in both cases.

Differences in the deuteron parameters among the models were
similarly quite small, with only the slight tendencies in
$d$-state probability, quadrupole moment and asymptotic $s$-state
predictions noted previously.

The motivation of this study was to see whether, with a very
restricted set of models, one type of coupling of the pseudoscalar
mesons would be able better to reproduce $NN$ scattering data than
the other.  Within the TOPT approach one might claim that, on the
whole, pseudovector coupling yields a slightly better description
of the data than pseudoscalar, but not in every partial wave.
Within the Thompson models, the differences between pseudovector
and pseudoscalar coupling are still smaller.  Between the TOPT and
Thompson model predictions there is no clear best.  We must
conclude, therefore, that the results obtained for $NN$ scattering
with the models considered in this work present no compelling
evidence that one form of coupling of the pseudoscalar mesons to
nucleons is favored over the other.

\vfill

\section*{Acknowledgments}
This work was performed in part under the auspices of the
U.~S. Department of Energy under contract No. DE-FG02-93ER40756 with
the Ohio University.

\pagebreak

\appendix

\section{Interaction Lagrangians and Hamiltonians}

We present here, for the purpose of completeness, the interaction Lagrangian
densities and the corresponding interaction Hamiltonian densities for the
various meson-nucleon interactions used in this work.

\subsection{Scalar meson}

\begin{equation}
{\cal L}_{I,s}=-g_s\overline{\psi}\psi\phi_s
\end{equation}
\begin{equation}
{\cal H}_{I,s}=-{\cal L}_{I,s}
\end{equation}

\subsection{ Pseudoscalar meson, pseudoscalar coupling}

\begin{equation}
{\cal L}_{I,pps}=-ig_p\overline{\psi}\gamma^5\psi\phi_p
\end{equation}
\begin{equation}
{\cal H}_{I,pps}=-{\cal L}_{I,pps}
\end{equation}

\subsection{Pseudoscalar meson, pseudovector coupling}

\begin{equation}
{\cal L}_{I,ppv}=-\frac{g_p}{2m_N}\overline{\psi} \gamma^5 \gamma^{\mu} \psi
\partial_{\mu} \phi_p
\end{equation}
\begin{equation}
{\cal H}_{I,ppv}=-{\cal L}_{I,ppv} +
\frac{1}{2} \frac{g_{p}^2}{4m_N^2}(\overline{\psi}\gamma^5 \gamma^0 \psi)^2
\end{equation}

\subsection{Vector meson}

Here we take for the free meson  Lagrangian the form

\begin{equation}
{\cal L}_{0,v}=-\frac{1}{4} F_{\mu \nu} F^{\mu \nu} \, + \,\frac{1}{2}m_v^2
A_{\mu} A^{\mu} ,
\end{equation}
where $F_{\mu \nu} \equiv \partial_{\mu} A_{\nu} \, - \, \partial_{\nu}
A_{\mu}$.  We then have

\begin{equation}
{\cal L}_{I,v} = -g_v\overline{\psi}  \gamma^{\mu} \psi A_{\mu} \,
- \,  \frac{f_v}{4m_N}\overline{\psi}  \sigma^{\mu \nu} \psi
F_{\mu \nu}
\end{equation}
\begin{equation}
{\cal H}_{I,v} = -{\cal L}_{I,v} + \frac{g_v^2}{2m_v^2}(\overline{\psi} \gamma ^0 \psi)^2
+\frac{1}{2} \frac{f_v^2}{4m_N^2}(\overline{\psi} \sigma^{i0} \psi)^2
\end{equation}
where $\sigma^{\mu \nu}  \equiv \frac{i}{2} [ \gamma^{\mu}, \, \gamma^{\nu}
]_{-}$.  We use the Bjorken-Drell conventions, summing on repeated indices and
using the Latin letter $i$ to denote spatial indices 1..3.  Isotopic spin notation
has been suppressed.

\section{Matrix elements of the potential}

We show here the potential matrix elements in time-ordered
perturbation theory to second order in the meson-nucleon coupling
constants derived from the Hamiltonians in Appendix A.   The
field-theoretic matrix element for the contribution of meson $j$
is multiplied by a cutoff function $ F_j(W,\vec{p}\,',\vec{p}\,)$,
as described in Section II.

We use the helicity basis in the two-nucleon center-of-mass (c.m.)
frame. The total c.m energy of the system is $W$, the momenta of
the ingoing and outgoing nucleons are $(\vec{p},\ -\vec{p}\,)$ and
$(\vec{p}\,',\ -\vec{p}\,')$ with corresponding energies $E_p$ and
$E_{p'}$ and helicities $\lambda_1,\ \lambda_2$ and $\lambda_1',\
\lambda_2'$, with $\lambda=\pm \frac{1}{2}$.  The energy transfer
in the interaction is $\delta = E_{p'}-E_p$, the three-momentum
transfer in the interaction is $\vec{k}=\vec{p}-\vec{p}\,'$ and
the energy of the exchanged meson of type $j$  is
$\omega_{k}^{j}=\sqrt{\vec{k}^2+m_j^2}$.  The energy denominator
for the  meson exchange term is given by $D^j \equiv
W-E_p-E_{p'}-\omega_{k}^j$.

\begin{widetext}
\subsection{Scalar meson}

\begin{equation}
V^{(2)}_s = \frac{ g_s^2}{(2\pi)^3}  {\rm I}_{12}\frac{1}{\omega_{k}^{s} D^s}
\left\{ [{\bf {1}}]_1 [{\bf {1}}]_2 \right\} F_s(W,\vec{p}\,',\vec{p}\,)
\end{equation}

\subsection{Pseudoscalar meson, pseudoscalar coupling}

\begin{equation}
V^{(2)}_{p(ps)} = -\frac{ g_p^2}{(2\pi)^3}
{\rm I}_{12}  \frac{1}{\omega_{k}^{p} D^s}
\left\{[\gamma^5]_1 [\gamma^5]_2 \right\}F_{p(ps)}(W,\vec{p}\,',\vec{p}\,)
\end{equation}

\subsection{Pseudoscalar meson, pseudovector coupling}

\begin{eqnarray}
V^{(2)}_{p(pv)} & = & \frac{ g_p^2}{(2\pi)^3}  {\rm I}_{12}
\bigg\{ \frac{1}{\omega_{k}^{p} D^p}
\bigg(
-[\gamma^5]_1 [\gamma^5]_2
-\frac{\delta}{2m_N}
\left( [\gamma^5 \gamma^0]_1[\gamma^5]_2+[\gamma^5]_1[\gamma^5 \gamma^0]_2 \right)
 \nonumber\\
& + & \frac{(\omega_k^p)^2-\delta^2}{4m_N^2}[\gamma^5 \gamma^0]_1[\gamma^5 \gamma^0]_2
\bigg)
-\frac{1}{4m_N^2}[\gamma^5 \gamma^0]_1[\gamma^5 \gamma^0]_2 \bigg\}
F_{p(pv)}(W,\vec{p}\,',\vec{p}\,)
\end{eqnarray}

\subsection{Vector meson}

\begin{eqnarray}
V^{(2)}_{v} & = &  \frac{g_v^2}{(2\pi)^3}  {\rm I}_{12} \bigg\{
\frac{1}{\omega_{k}^{v} D^v} \bigg( - g_{\mu \nu} [\gamma^{\mu}]_1
[\gamma^{\nu}]_2
+\frac{(\omega_{k}^{v})^2-\delta^2}{m_v^2}[\gamma^0]_1[\gamma^0]_2
\bigg) \nonumber\\
& + & \frac{1}{m_v^2}{[\gamma^0]_1[\gamma^0]_2}
\bigg\}  F_{v}(W,\vec{p}',\vec{p}) \nonumber\\
& + & \frac{{f}_v^2}{(2\pi)^{3}4m_N^2}  {\rm I}_{12} \bigg\{
\frac{1}{\omega_{k}^{v} D^v} \bigg(
 g_{\mu \nu}
[\sigma^{\alpha\mu}(p_1'-p_1)_\alpha]_1
[\sigma^{\beta\nu}(p_2'-p_2)_\beta]_2
\nonumber\\
& + & \delta \Big( [\sigma^{0i}]_1[\sigma^{\beta
i}(p_2'-p_2)_{\beta}]_2 +[\sigma^{\alpha
i}(p_1'-p_1)_{\alpha}]_2[\sigma^{0i}]_2 \Big)
+((\omega_{k}^{v})^2-\delta^2)[\sigma^{0i}]_1[\sigma^{0i}]_2 \bigg) \nonumber\\
& + & [\sigma^{0i}]_1[\sigma^{0i}]_2 \bigg\}
F_{v}(W,\vec{p}',\vec{p}) \nonumber\\
& + & \frac{{g}_v {f}_{v}}{(2\pi)^{3}2m_N}  {\rm I}_{12}
\frac{1}{\omega_{k}^{v} D^v} \bigg\{ g_{\mu \nu}([\gamma^{\mu}]_1
[\sigma^{\alpha\nu}(p_2'-p_2)_\alpha]_2+
[\sigma^{\alpha\mu}(p_1'-p_1)_\alpha]_1[\gamma^{\nu}]_2)           \nonumber\\
& + &
\delta([\gamma^{i}]_1[\sigma^{0i}]_2+[\sigma^{0i}]_1[\gamma^{i}]_2)
\bigg\} F_{v}(W,\vec{p}',\vec{p})
\end{eqnarray}

In the expressions above we have used the compressed notation

\begin{equation}
V^{(2)}_{j} \equiv  <\vec{p}\ '\lambda_1'\lambda_2'\mid V^{(2)}_{j}(W)                        \mid\vec{p}\lambda_1\lambda_2>
\end{equation}
and
\begin{equation}
[{\mathbf A}]_1[{\mathbf B}]_2 \equiv
\left[\overline{u}(\vec{p}\ ',\lambda_{1}'){\mathbf A}u(\vec{p},\lambda_{1})\right]
\left[\overline{u}(-\vec{p}\ ',\lambda_{2}'){\mathbf B}u(-\vec{p},\lambda_{2})\right]
\end{equation}
\end{widetext}
for the matrix elements of the potential.

The isospin factor ${\rm I}_{12}$ is 1 or $\tilde{\tau}_1 \cdot \tilde{\tau}_2$
as the isospin of the meson concerned is 0 or 1. The normalization of the Dirac
spinors is
\begin{equation}
u^{\dag}(\vec{p},\lambda)u(\vec{p},\lambda)=1,
\end{equation}
and the four-vectors $p_1$ and $p_2$ are $(E_p,\vec{p}\,)$ and $(E_p,-\vec{p}\,)$,
respectively, and similarly for the primed quantities.

We have written the matrix elements in a way to distinguish
clearly the meson exchange terms, which contain energy
denominators $D^j=W-E_{p'}-E_{p}-\omega_{k}^{j}$, from the contact
terms, which do not. The equivalence of the fully on-shell
one-pion exchange with pseudovector coupling with that of the one
with pseudoscalar coupling is then evident. In that case
$E_{p}=E_{p'}=W/2$, $D^p=-\omega_{k}^{p}$ and $\delta=0$.  The
term containing $(\omega_{k}^{p})^2$ is exactly canceled by the
contact term.

\section{The Thompson equation}

Matrix elements for the potential to be used in the Thompson equation
can easily be found using the results of Appendix B by applying the
following prescription: \newline \newline \noindent
1. Change the nucleon spinor normalization to
\begin{equation}
\overline{u}(\vec{p},\lambda)u(\vec{p},\lambda)=1.
\end{equation}
\newline \newline \noindent
2. For each meson contribution  $V^{(2)}_j$, replace the energy
denominator $\omega^{j}_{k}D^j$ according to
\begin{equation}
\omega^{j}_{k}D^j \longrightarrow -(\omega^{j}_{k})^2 \ = \
-(\vec{p}\ '-\vec{p}\,)^2-m_j^2
\end{equation}
\newline \newline \noindent
3. Drop all contact terms.

\bibliography{pspv}

\newpage

\begin{table}
\caption{\label{props} Quantum numbers and masses of the mesons used
in the models in this work.}
\begin{ruledtabular}
\begin{tabular}{ccc}
meson & ${\rm I~(J^P)}$ & mass (GeV) \\
\hline
$\pi$ & $1~(0^-)$ & 0.13803 \\
$\eta$ & $0~(0^-)$ & 0.5488 \\
$\sigma$ & $0~(0^+)$ & 0.52 \\
$a_0$ & $1~(0^+)$ & 0.983 \\
$\rho$ & $1~(1^-)$ & 0.769 \\
$\omega$ & $0~(0^-)$ & 0.7826 \\
\end{tabular}
\end{ruledtabular}

\end{table}

\begin{table}
\caption{\label{params} Parameters for the TOPT models.  The
columns labelled PV and PS represent, respectively, the values for
the model with pseudovector and pseudoscalar coupling of the $\pi$
and $\eta$ mesons, as in the figures.  Fixed parameters are shown
in parentheses. The values of searched parameters are determined
by a least-squares fit to the SP40 $NN$ phase shift analysis of
Ref.~\onlinecite{said}.  For these models $n_\pi=1$ (see
Eq.~(\ref{formfac})).}
\begin{ruledtabular}
\begin{tabular}{cccccc}
Coupling constant& PV & PS & Cutoff & PV(GeV) & PS(GeV) \\
\hline
$g_{\pi}^2/4\pi$ & (13.8) & (13.8) & $\Lambda_\pi$ & 2.50 & 1.80 \\
$g_{\eta}^2/4\pi$ & 2.15 & 2.00 & $\Lambda_\eta$ & 1.00 & 1.21 \\
$g_{\sigma}^2/4\pi$ & 6.443 & 7.177 & $\Lambda_\sigma$ & 10.00 & 9.80 \\
$g_{a_0}^2/4\pi$ & 0.858 & 1.859 & $\Lambda_{a_0}$ & 2.50 & 1.50 \\
$g_{\rho}^2/4\pi$ & 1.14 & 1.06 & $\Lambda_\rho$ & 1.45 & 1.73 \\
$f_{\rho}/g_{\rho}$ & 5.12  & 4.40 &  & & \\
$g_{\omega}^2/4\pi$ & 17.40 & 22.20 & $\Lambda_\omega$ & 1.41 & 1.35 \\
$f_{\omega}/g_{\omega}$ & (0) & (0) &  & & \\
& & & $\Lambda_N$\footnote[1]{See Section~II, Eq.~(5) for the use
of $\Lambda_{N}$.} & (0.7) & (0.7) \\
\end{tabular}
\end{ruledtabular}
\end{table}

\begin{table}
\caption{\label{tparams}  Parameters for the Thompson equation
models. The columns labelled TPV and TPS represent, respectively,
the values for the model with pseudovector and pseudoscalar
coupling of the $\pi$ and $\eta$ mesons, as in the figures.  Fixed
parameters are shown in parentheses.  The values of searched
parameters are determined by a least-squares fit to the SP40 $NN$
phase shift analysis of Ref.~\onlinecite{said}.}
\begin{ruledtabular}
\begin{tabular}{cccccc}
Coupling constant& TPV & TPS & Cutoff & TPV(GeV) & TPS(GeV) \\
\hline
$g_{\pi}^2/4\pi$ & (13.8) & (13.8) & $\Lambda_\pi$ & 2.00 & 1.80 \\
$g_{\eta}^2/4\pi$ & 4.22 & 5.00 & $\Lambda_\eta$ & 1.00 & 1.00 \\
$g_{\sigma}^2/4\pi$ & 6.871 & 7.205 & $\Lambda_\sigma$ & 10.00 & 10.00 \\
$g_{a_0}^2/4\pi$ & 5.00 & 5.00 & $\Lambda_{a_0}$ & 2.50 & 1.50 \\
$g_{\rho}^2/4\pi$ & 0.800 & 0.800 & $\Lambda_\rho$ & 1.34 & 1.31 \\
$f_{\rho}/g_{\rho}$ & 6.89  & 6.89 &  & & \\
$g_{\omega}^2/4\pi$ & 25.00 & 25.00 & $\Lambda_\omega$ & 1.245 & 1.260\\
$f_{\omega}/g_{\omega}$ & (0) & (0) &  & & \\
& & & $\Lambda_N$ & (0.7) & (0.7) \\
\end{tabular}
\end{ruledtabular}
\end{table}

\begin{table}
\caption{\label{deuteron}  Deuteron properties calculated with the
four models considered.
 }
\begin{ruledtabular}
\begin{tabular}{cccccc}
Quantity & Experiment & PV & PS & TPV & TPS \\
\hline
$-E_d$~[MeV] & 2.24575(9) \cite{vdLeun} & 2.22450 & 2.22447 & 2.22453 & 2.22459 \\
$P_D$ [\%] &  & 3.8 & 3.4 & 4.4 & 3.9 \\
$Q_d$~[$\rm fm^2$]\footnote[1]{Theoretical values do not include
meson exchange current
contributions.} & 0.2859(3) \cite{Eric,Bish} & 0.2784 & 0.2780 & 0.2765 & 0.2754 \\
$A_S$ [fm$^{-1/2}$] & 0.8846(9) \cite{Eric,STS} & 0.9117 & 0.9119 & 0.8974 & 0.8950 \\
$A_D/A_S$ & 0.0256(4) \cite{Rodning} & 0.0255 & 0.0260 & 0.0252 & 0.0257 \\
\end{tabular}
\end{ruledtabular}
\end{table}

\clearpage 


\begin{figure}[T]
\includegraphics{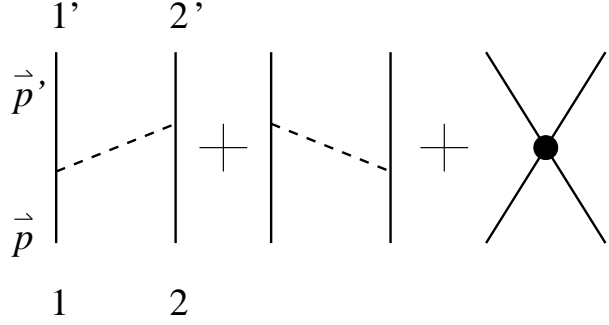} 
\caption{\label{fig:topt} Schematically, the three terms in
time-ordered perturbation theory that may contribute to $NN$
scattering in second order of the coupling constant in a meson
exchange model.}
\end{figure}
\begin{figure}[t]
\vspace{5mm}
\includegraphics{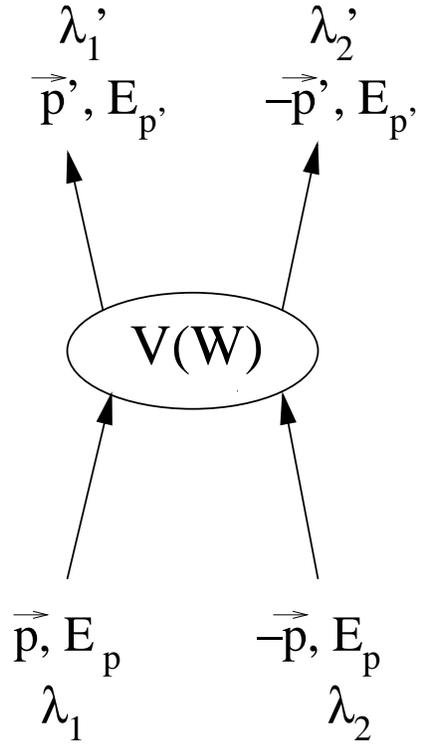}
\caption{\label{fig:potential}. Kinematics for the $NN$ potential
in the $NN$ center- of-mass frame with total energy
\textit{\textbf{W}}. The fermion lines are labelled by their
3-momentum, energy and helicity.}
\end{figure}
\begin{figure}
\psfig{file=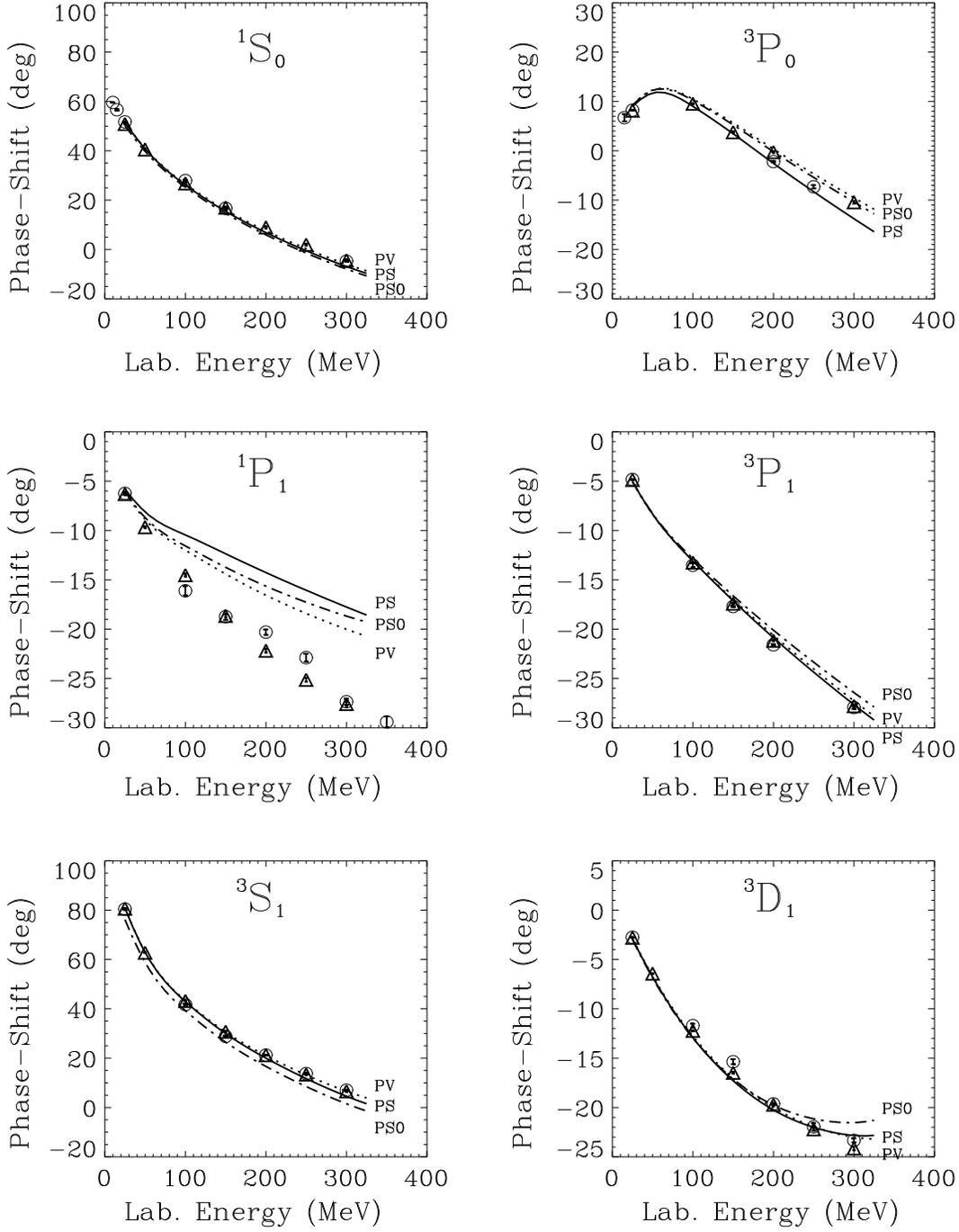,width=15.0cm,height=19.0cm}
\caption{\label{fig:toptphases}. Phase shifts for TOPT models. The
curves labelled PV are the results of the best fit for the model
with pseudovector coupling of the $\pi$ and the $\eta$; those
labelled PS are the best fit for pseudoscalar coupling.  The
curves labelled PS0 are the results of using the parameter set of
PV but with pseudoscalar coupling.  Parameters for the models are
given in Table~\ref{params}. The triangles represent the phase
shift analysis by the Nijmegen Group at selected energies
\cite{nnonline}, and the open circles stand for the energy
independent analysis SP40 from the CNS DAC Services SAID
\cite{said}.}
\end{figure}

\begin{figure}
\psfig{file=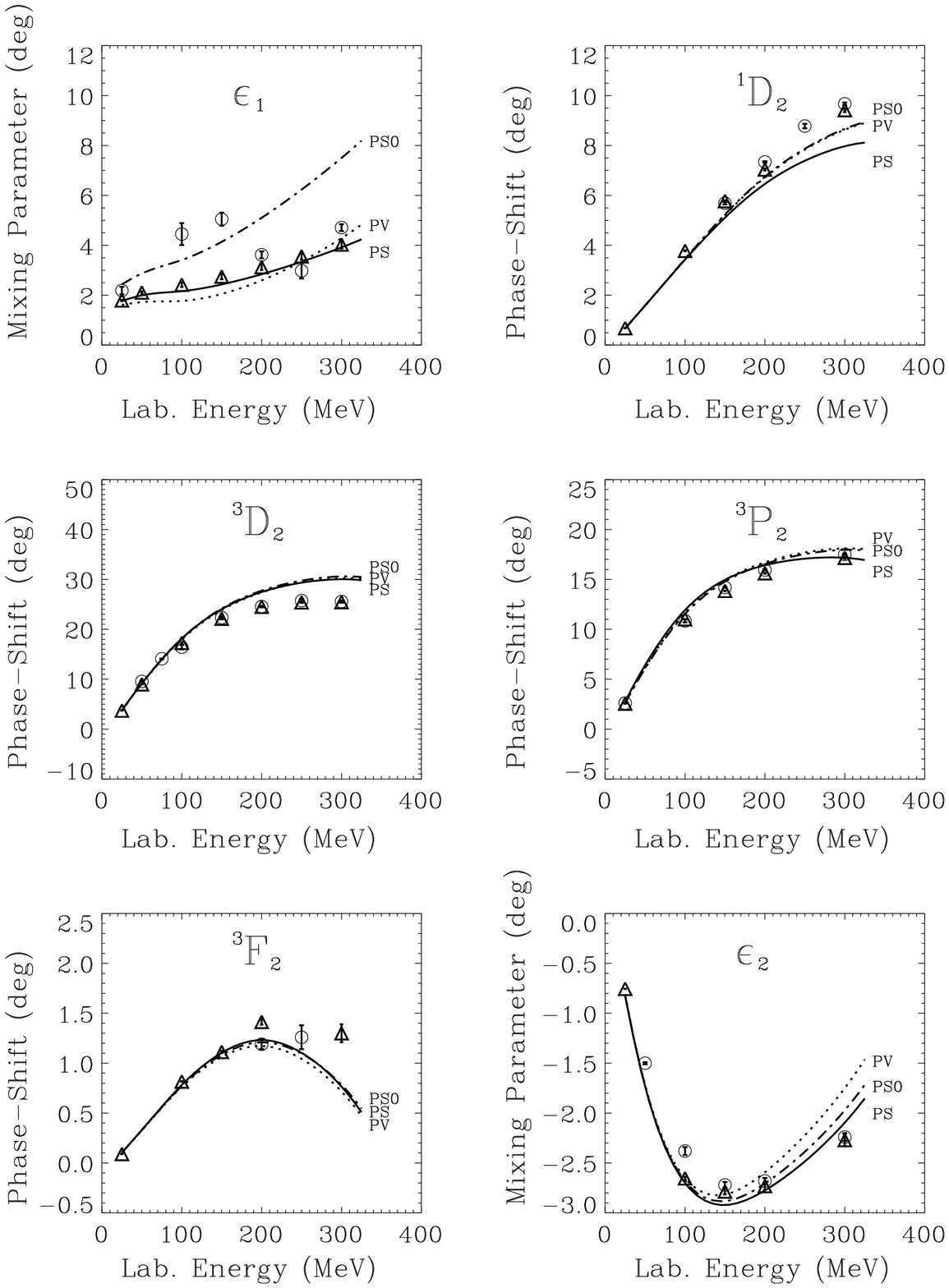,width=15.0cm,height=19.0cm}
\centerline{Fig.
3 cont'd}
\end{figure}
\begin{figure}
\psfig{file=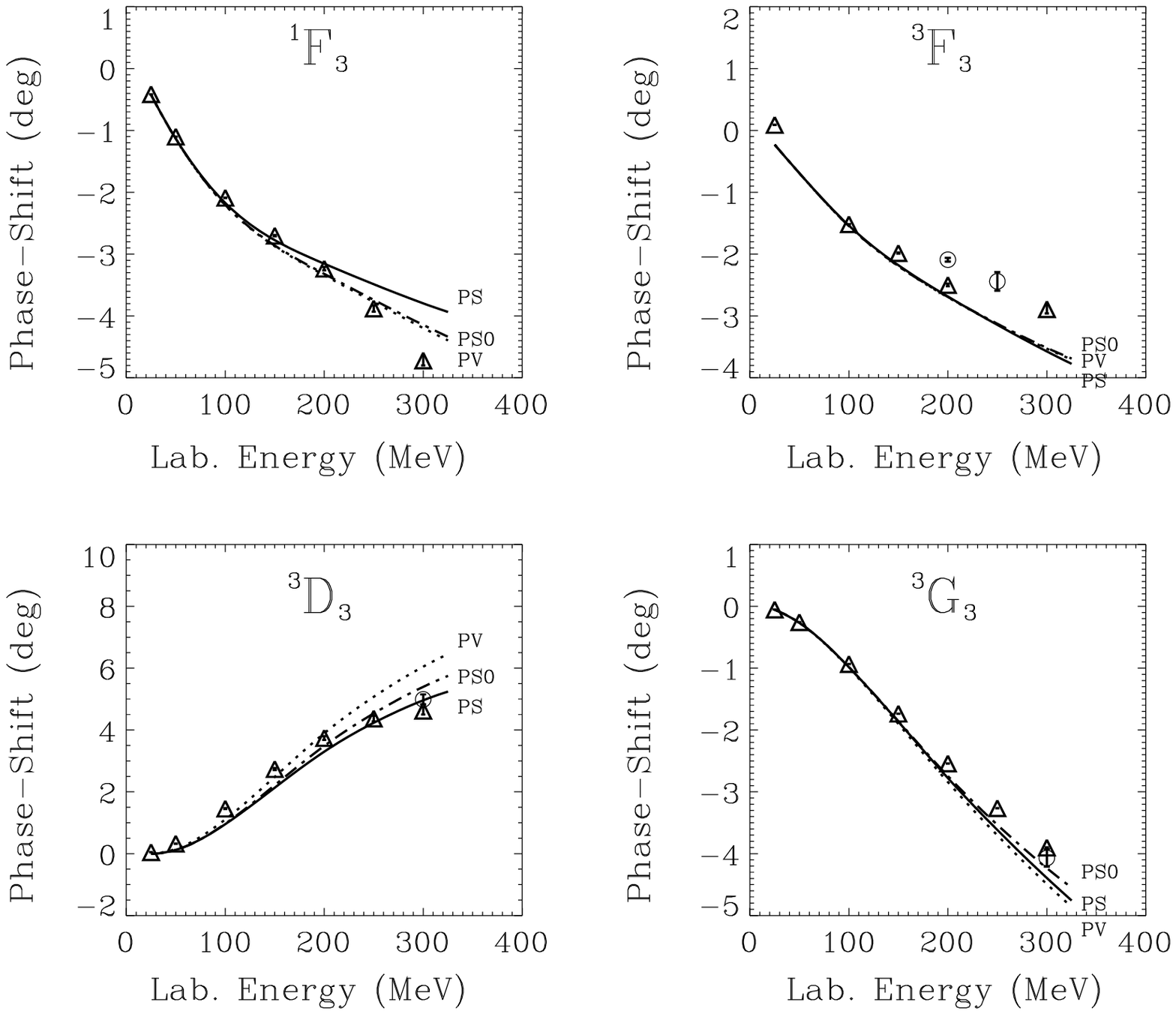,width=15.0cm,height=19.0cm} \centerline{Fig.
3 cont'd}
\end{figure}
\begin{figure}
\psfig{file=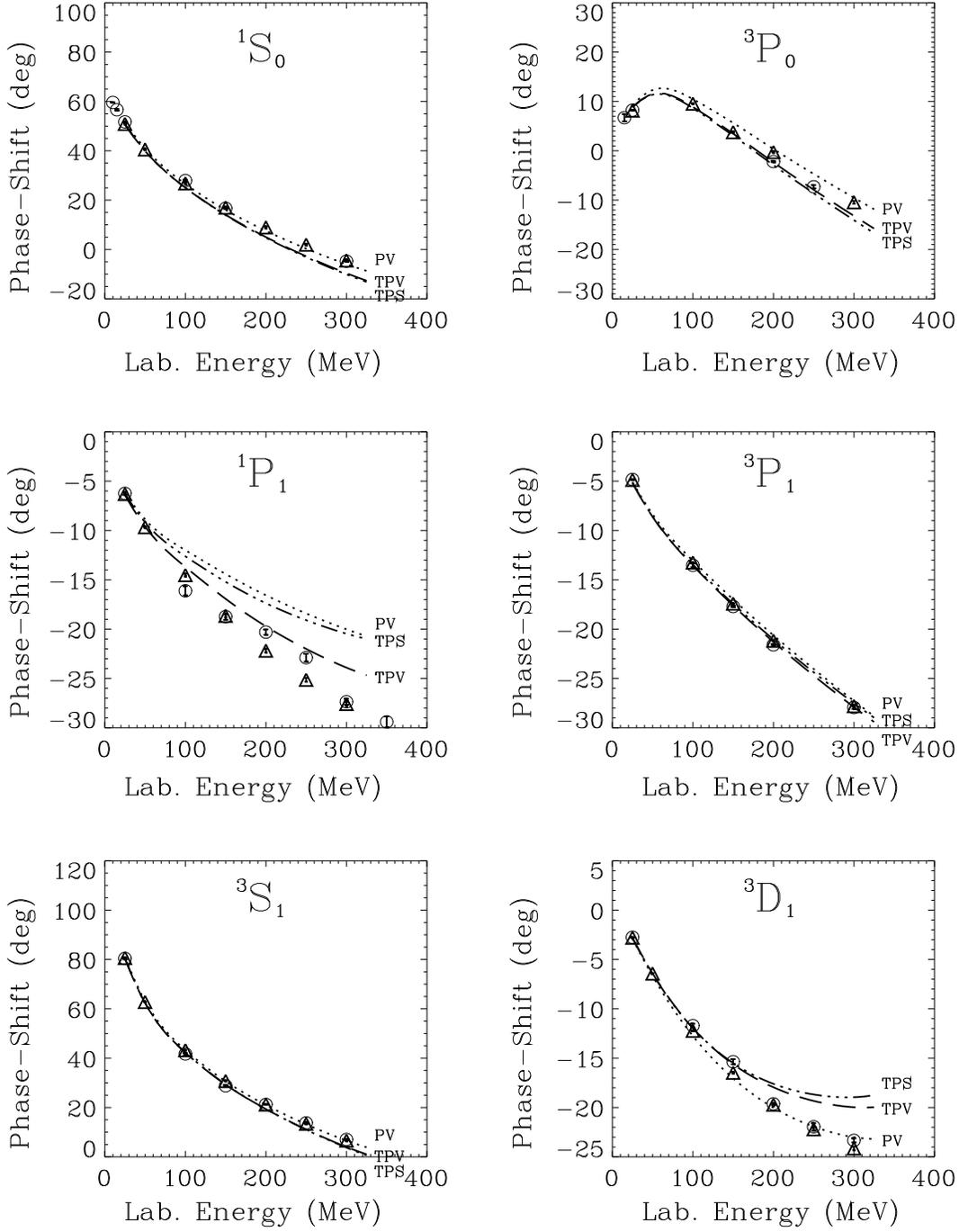,width=15.0cm,height=19.0cm}
\caption{\label{fig:thompsonphases}. Phase shifts for Thompson
equation models. The curves labelled TPV are the results of the
best fit for the model with pseudovector coupling of the $\pi$ and
the $\eta$; those labelled TPS are the best fit for pseudoscalar
coupling. Included for purposes of comparison are the results of
the best fit of the TOPT model with pseudovector coupling (curves
labelled PV). Parameters for the models are given in
Table~\ref{params}. The error bars are the same as in Fig.~3}
\end{figure}

\begin{figure}
\psfig{file=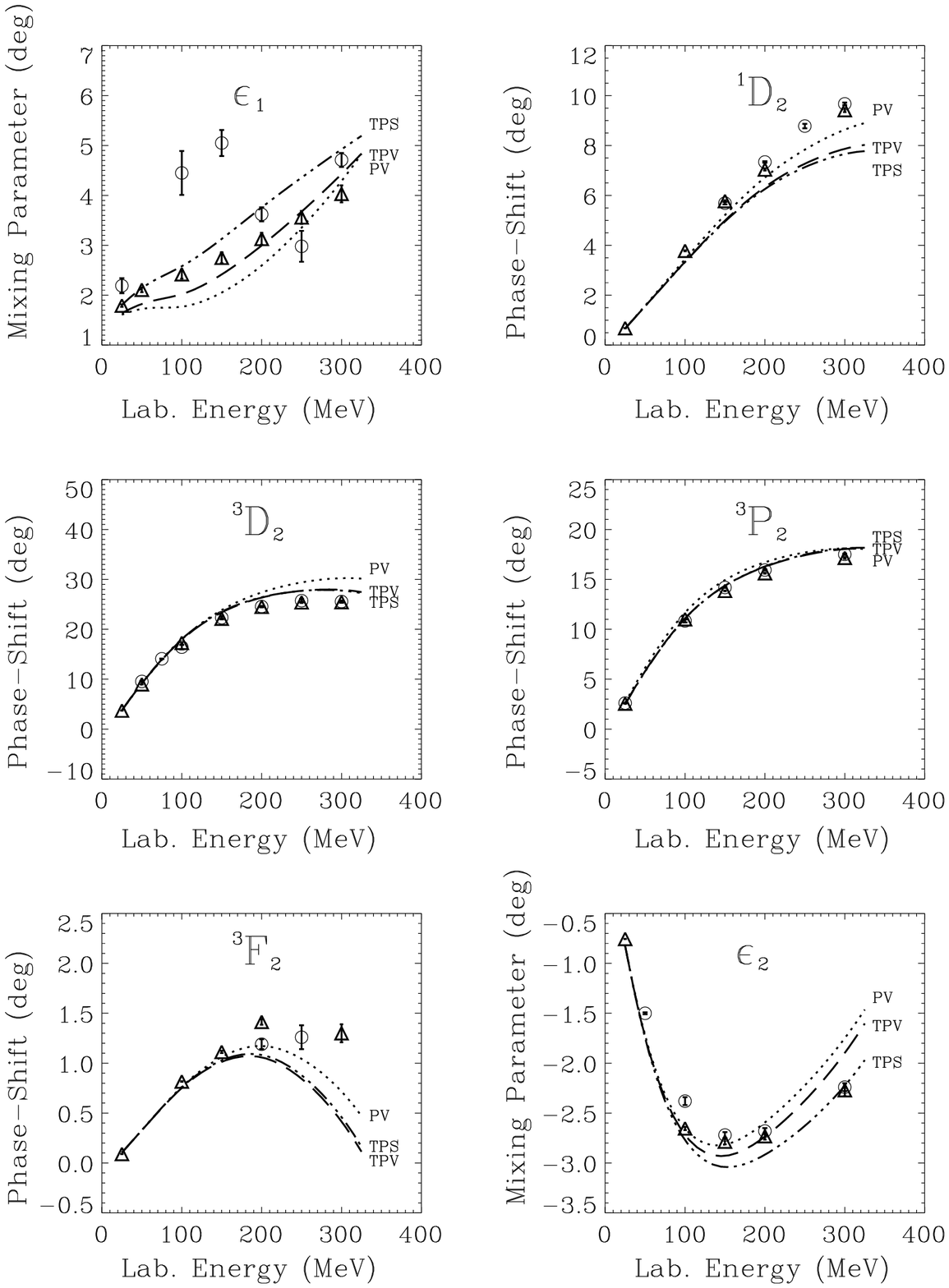,width=15.0cm,height=19.0cm}
\centerline{Fig. 4 cont'd}
\end{figure}
\begin{figure}
\psfig{file=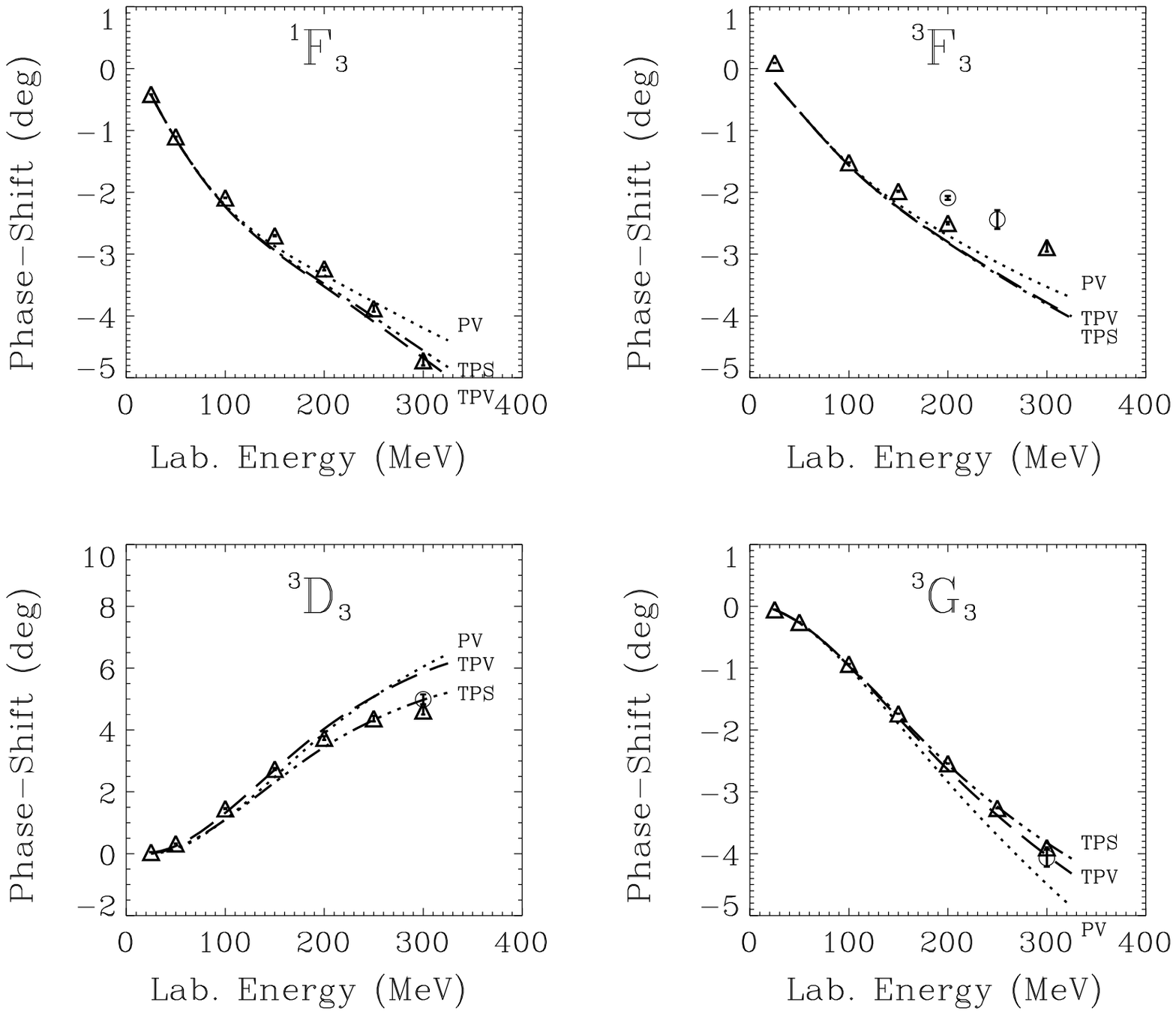,width=15.0cm,height=19.0cm}
\centerline{Fig. 4 cont'd}
\end{figure}
\end{document}